\documentclass[aps,prl,twocolumn,groupedaddress,showpacs]{revtex4}
\usepackage{amsmath}
\usepackage{amssymb}
\usepackage{graphicx}
\usepackage{color} 

\newcommand{\vect}[1]{\boldsymbol{#1}}

\begin{document}

\title{Cancer Metastasis: Collective Invasion in Heterogeneous Multicellular Systems}
\author{Adrien Hallou}
\affiliation{Department of Engineering, University of Cambridge, Cambridge, CB2 1PZ,UK.}
\author{Joel Jennings}
\affiliation{Department of Engineering, University of Cambridge, Cambridge, CB2 1PZ,UK.}
\author{Alexandre Kabla}
\email{ajk61@cam.ac.uk}
\affiliation{Department of Engineering, University of Cambridge, Cambridge, CB2 1PZ,UK.}

\date{\today}

\begin{abstract}
Heterogeneity within tumour cell populations is associated with an increase in malignancy and appears to play an important role during cancer metastasis. 
Using \textit{in silico} experiments, we study the interplay between collective behaviours and cell motility heterogeneities in a model system.
Working with tumour spheroids that contain two non-proliferating cell populations of different motile properties, we explore the conditions required for maximal invasion into surrounding tissues. 
We show emerging spatial patterns of cellular organisation and invasion which are consistent with \textit{in vitro} and \textit{in vivo} observations. 
This demonstrates that mechanical interactions at the cellular level are sufficient to account for many of the observed morphologies of invasion and that heterogeneity in cell motility can be more important than average mechanical properties in controlling the fate of large cell populations. 
\end{abstract}


\maketitle

Metastasis, the process during which cancer cells migrate away from primary tumour and disseminate in other organs, accounts for more than $90\%$ of cancer fatalities \cite{Wirtz2011}. 
At the cellular level, malignant cell behaviour has been associated with an accumulation of gene mutations and a malfunction of numerous regulatory signalling pathways \cite{Hanahan2000}.
Moreover, following earlier pioneering works \cite{Fidler1978}, recent analyses have established that most malignant solid tumours are a mosaic of different cell populations with different phenotypic traits \cite{Marusyk2012}.\
Owing to the variability in these factors across different pathologies, patients and even tumours, no common mechanistic pathway leading to the induction and progression of metastasis has been yet characterized \cite{Vanharanta2013,Wan2013}.\\
In contrast, recent advances in imaging techniques at the tissue scale  \cite{Alexander2008,Weigelin2012,Friedl2012}, along with histopathological studies on patients' biopsies \cite{Friedl2012}, draw a surprisingly unified picture of cancer invasion in terms of cell migration behaviours.
In most cancers, despite tremendous variations in gene expression and biochemical environment, tumour cells are able to invade collectively by maintaining intercellular coordination \cite{Friedl2003,Friedl2012,Gaggioli2007}, generating hence multicellular structures such as cell sheets, strands or clusters remaining cohesive and polarised \cite{Friedl2003}.\
Accumulating evidence suggests that the genetic and phenotypic heterogeneity observed in these structures plays a key role in the invasion of the tumour cells into surrounding tissues. Collective invasion of otherwise non-invasive tumour cells can be driven by invasion-competent cancer cells \cite{Cheung2013,Carey2013,Shin2014,Chapman2014} or stromal fibroblasts \cite{Dang2011,Gaggioli2007} through a ``leader / follower" invasion mechanism \cite{Friedl2012}.\
These patterns of collective behaviour suggest that a system-level analysis is needed to account for the generic aspects of tumour cell invasion.\\ Cooperative behaviours are observed in a number of biological contexts such as collective animal behaviours \cite{Deisboeck2009}, glassy dynamics of epithelial tissues \cite{Mark2010,Trepat2011,Angelini2011,Tambe2011}, wound healing \cite{Poujade2007,Anon2012,Cochet-Escartin2014} and early developmental processes \cite{Friedl2009a, CarmonaFontaine2008, Dumortier2012}. Moreover, a number of physical factors modulating collective behaviours in tissues have also been investigated: geometrical confinement of cells \cite{Vedula2012,Doxzen2013,Deforet2014}, mechanical and topographical properties of the extracellular matrix \cite{Pathak2007,Zaman2006,Londono2014} and forces applied by neighbouring tissues \cite{Basan2011,Montel2011,Tse2012,Delarue2013}.\
Physical models inspired by self-propelled particles \cite{Vicsek1995,Gregoire2004} have recapitulated some of these experimental findings and demonstrated that interactions between cell populations and their environments can be understood without resorting to subcellular processes \cite{Sepulveda2013,Basan2013}.\
But most of these studies have focused so far on well controlled homogeneous cell populations and the role of heterogeneity in cell populations remains poorly addressed, both experimentally and theoretically.\\
In this letter, we use the framework introduced in \cite{Kabla2012} to analyse the effect of cell heterogeneity on tumour invasion.
The model involves a Cellular Potts algorithm \cite{Graner1992} including where appropriate a self-propelled term to account for the active motion of the cells. 
Tumours are modelled as 2D packed populations of motile cells of spheroid shape surrounded by cohesive tissues made of passive cells that mechanically resist invasion. 
Although real tumours are far more complex, this approach provides a simple dynamic framework to analyse in general terms the interplay between cell motility and cell-cell interactions in cell invasion.
We will first study the initiation of invasion in the case of homogeneous tumours as a benchmark case. 
Timescales involved in this study are short enough to ignore tumour growth and cell proliferation. 
We then quantify the effect of population heterogeneity by introducing a small proportion of highly motile cells into the system.\\
\begin{figure}[!ht]
\begin{center}
\includegraphics[width=3in]{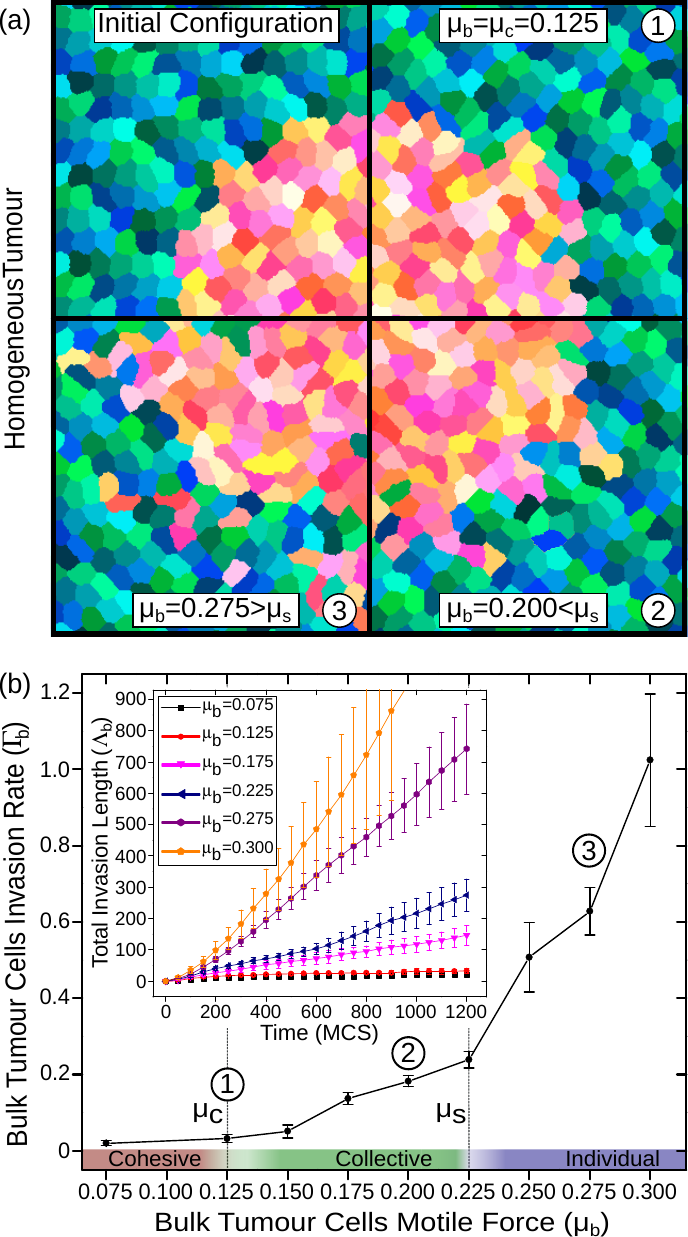}
\end{center}
\vspace{-5mm}
\caption{Collective invasion for homogeneous tumours. a) Snapshots of \textit{in silico} experiments. Top-left quadrant represents the initial state of the system (green-blue cells are non-motile, yellow-red cells are bulk tumour cells). The three other quadrants represent the state of the system for different values of $\mu_b$ at $t=4000$ MCS. b) $\Gamma_b=f(\mu_b)$. Each data point is evaluated for a given $\mu_b$ as the mean gradient of $\Lambda_b=f(\mu_b)$ between $0$ and $1200$ MCS. Inset: Total invasion length of bulk tumour cells $\Lambda_b$ in respect with time, for different values of $\mu_b$. Error bars are standard deviations.\label{Fig1}}
\vspace{-10pt}
\end{figure}
In the model, tissues are represented on a 2D lattice of $1280$ by $1280$ pixels, where the value $i$ of each pixel codes for the identity of the cell on that location. 
The cell volume $V_i(t)$ is constrained to a target value $V_0$ $[L^2]$ by a bulk modulus $\kappa$ $[E/L^{2}]$. 
The interfacial cell-cell interactions, such as adhesion or cortical tensions, are all accounted for by a single parameter, $J$ $[E/L]$, which sets the surface energy of cell membrane. 
The active motion of each cell is driven by a motile force $\vect{f}_{i} \left(t\right) =  \mu_{i} \vect{n}_{i}$ acting on a fixed substrate, where $\vect{n}_{i}\left(t\right)$ defines the cell polarization axis and $\mu_{i}$ $[E/L]$ its active motile force magnitude, which depends on the cell type.
The motile force is used to calculate a migration energy function which captures the mechanical work generated by each active cell:  $w_i = -\vect{f}_i \cdot \vect{r}_{i} $ $[E]$,  where $\vect{r}_{i}$ represents the position of the cell centroid. At any given time, volume constraint, cell-cell interactions and active migration energy terms can be combined in to an overall energy function, $E (t)$:
\begin{equation}
\begin{aligned}
E &= \sum_{k,k'} J \left( 1 - \delta_{i(k) i(k')}\right) + \sum_{i} \left( \frac{\kappa}{2} \left( V_{i}-V_0 \right)^2 + w_i \right)
\end{aligned}
\end{equation}
where  $i$  represents the cell index and $(k,k')$ represent pairs of neighbouring pixels. $\delta_{i(k)i(k')}$ is 1 when both pixels belong to the same cell and $0$ otherwise.
The system dynamics results from the iterative minimisation of this energy function through the Metropolis Monte-Carlo algorithm \cite{Kabla2012}. 
Time is here expressed in Monte Carlo steps (MCS), where 1 MCS corresponds to an average of one iteration per pixel over the whole lattice. 
The polarization vector of a cell, $\vect{n}_{i}(t)$, also evolves over time and is set along the direction of the mean previous displacements of the cell in the time interval $[t-\tau,t]$; $\tau$ represents the time-scale at which cell polarity responds to external stimuli and controls the persistence length of the cell trajectory in the absence of other cells \cite{Szabo2010}.
To highlight the role of cell migration, all cell populations have the same mechanical, interfacial and dynamical properties ($V_0 =400$ $pixel^2$, $\kappa=1$, $J=5$, $\tau=10$ MCS, noise level $T=2.5$ [E]). They only differ in the value taken by their motile forces. Earlier work \cite{Kabla2012} has studied in detail the role of the different model parameters and identified two important behavioural transitions in the populations dynamics (\textit{cf.} the Table 1 of supplementary materials): 
(i) A transition from static epithelial behaviour to collective streaming in homogeneous motile populations, at a critical value of the motile force parameter $\mu_c \approx 0.125$. 
(ii) A critical force $\mu_s \approx 0.225$ needed for a single cell to migrate through a non motile population. 

We first quantify the short time scale dynamics of invasion in homogeneous tumour spheroids. These structures consist of bulk tumour cells of motile force $\mu_b$ and possess an initial radius $R_0$ of the order of 10 cell diameters corresponding to a population of around $300$ cells. 
Invasion is measured as a cumulative length $\Lambda_b(\mu_b,t)$ which is the sum of the distances to the initial tumour boundary for all cancer cells ${i}$ who left the tumour.
This averaged quantity encompasses both the spatial extent of invasion and the number of invading cells. 
$\Lambda_b$ varies almost linearly with time for the first thousand MCS as shown in the inset of Fig.~\ref{Fig1}.b.
The degree of malignancy of the system, \textit{i.e} the capacity of bulk tumours cells to invade surrounding tissues, can be characterised as a rate of invasion at short time scales, $\Gamma_b(\mu_b) =  \Lambda_b(\mu_b,\Delta t)/\Delta t$. 
The evolution of $\Gamma_b$  as a function of the cells' motile force $\mu_b$ is represented on figure Fig.\ref{Fig1}.b, for $\Delta t = 1200$ MCS.
Homogeneous tumours display behaviours ranging from no invasion to weak collective protrusions to widespread single cell dispersal (see Fig.1.a). Transitions are controlled by the two critical motile forces $\mu_c$ and $\mu_s$ previously defined (see movies 1 and 2 in supplementary materials). 
These qualitative transitions are reflected in the invasion rate as shown in Fig.1.b. 
The regime where $\mu_c < \mu_b < \mu_s$ is remarkable as invasion, although slow, is observed and takes the form of finger-like collective patterns (\textit{cf.} Fig.\ref{Fig1}.a bottom-right quadrant and movie 3 in supplementary materials), as described in many \textit{in vivo} studies \cite{Alexander2008,Weigelin2012,Friedl2012}.
This suggests that the emergence of finger-like protrusions is the result of simple physical interactions in a regime where single cells are not strong enough to invade the surrounding tissue, but coordinated cell groups can do so by collectively generating larger forces. 
This process occurs without any directional cue or specific interaction between cells, other than mechanical forces between cells.
While this might capture a fundamental mechanism involved in the early stages of collective invasion, the rates are however low compared to what is qualitatively expected from the experimental literature \cite{Cheung2013} and would not be sufficient to explain why these are associated with highly malignant situations.
The sensitivity and convexity of the invasion rate to the motile force of the tumour cells suggests that introducing heterogeneities in the motile properties of the cancer cells could have a significant effect on invasion rates and patterns.\\
\begin{figure}[!ht]
\begin{center}
\includegraphics[width=3in]{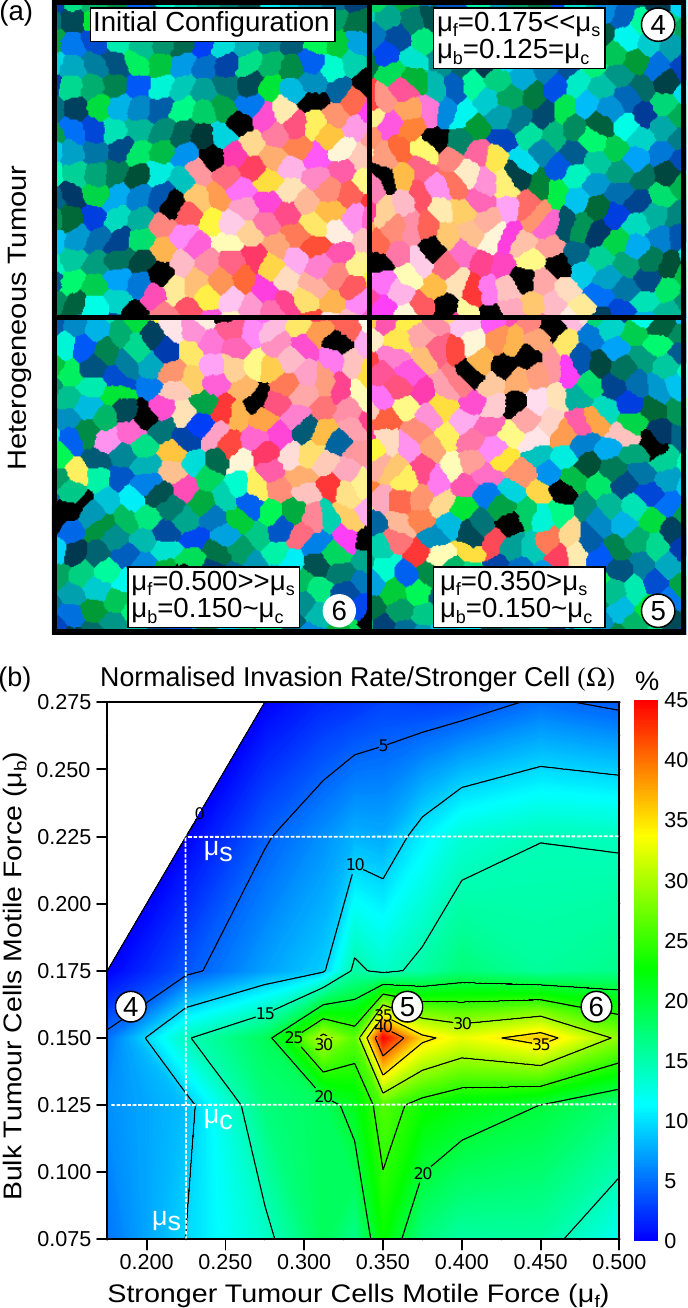}
\end{center}
\vspace{-5mm}
\caption{a) Snapshots of \textit{in silico} experiments with $24$ stronger cells. Top-left quadrant represents the initial state of the system (green-blue cells are non-motile, yellow-red cells are bulk tumour cells, black cells are stronger tumour cells). The three other quadrants represent the states of the system for different values of $\mu_b$ and $\mu_{f}$ at $t=4000$ MCS. b) Heat map of the normalized invasion rate per stronger cell $\Omega$. The maximum is observed for $\mu_b=0.150$ and $\mu_{f}=0.350$. Data set is made of $58$ points averaged each over $12$ different simulations. \label{Fig2}}
\vspace{-10pt}
\end{figure}
We now consider a tumour spheroid surrounded by a small number $N_f$ of  cells with a larger motile force ($\mu_f$), inspired by fibroblasts or stronger tumour cells sub-populations but representing more generally a certain level of heterogeneity in the system (\textit{cf.} Fig.~\ref{Fig2}.a). 
We analysed the system's response for $N_f=$ 6, 12 or 24, i.e. about 2\%, 4\% or 8\% of the total motile cell population, respectively. 
The invasive behaviour of bulk tumour cells in the presence of these particular cells is quantified as previously by the invasion length $\Lambda_f(\mu_b,\mu_f,N_f,t)$ and the corresponding invasion rate $\Gamma_f(\mu_b,\mu_f,N_f)= \Lambda_f(t)/\Delta t$.
To quantify more specifically the effect of heterogeneity on the invasion efficiency, we extract the percentage increase in the invasion rate per stronger cell: $\Omega(\mu_b,\mu_f)= \left(\partial_{N_f} \Gamma_f \right) / \Gamma_b$ (Fig.~\ref{Fig2}.b).
As expected, the presence of stronger cells brings no significant enhancement to the invasion efficiency for $\mu_b \geq \mu_s$, where the bulk of the cancer cells are able to invade on their own (see movie 4 in supplementary materials).
When both bulk and stronger tumour cells generate a motile force below the threshold for single cell invasion, there is only a marginal effect on the invasion (see movie 5 in supplementary materials). 
In contrast, when the bulk of the tumour is in the regime of collective invasion $(\mu_c \leq  \mu_b \leq \mu_s)$, the addition of a small number of stronger cells  $(\mu_f \geq \mu_s)$ has a dramatic effect. Each of them increases significantly the invasion rate $(20 \% \leq  \Omega \leq 45 \%)$ for a restricted $\mu_f$ value range. 
The increase in invasion rate has an optimum for $\mu_b\approx 0.150$ and $\mu_f \approx 0.350$. To understand the significance of this optimum, we analysed the morphology of the tumour at the onset of invasion.\
\begin{figure}[!ht]
\begin{center}
\includegraphics[width=3in]{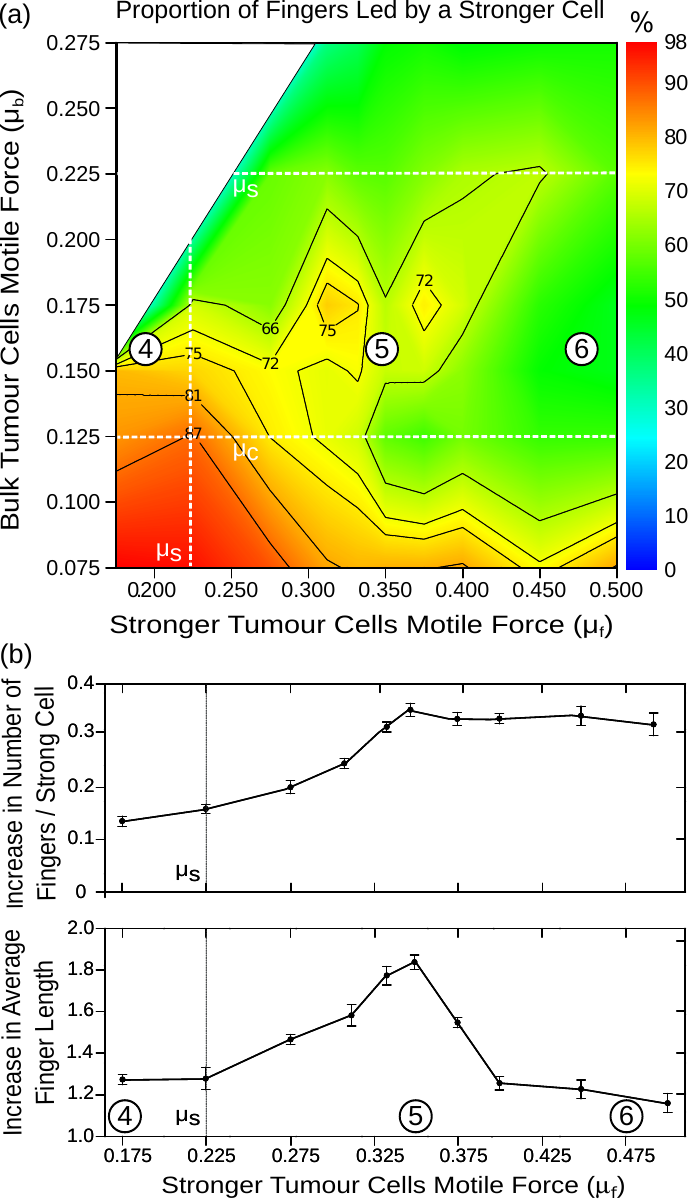}
\end{center}
\vspace{-5mm}
\caption{Morphology of invading structures in heterogeneous tumours. a) Heat map of the proportion of fingers led by a stronger cell. Measurements are taken on five time points between 3500 and 4500 MCS, and averaged for each on 12 different seeds. Data set is the same as for Fig.~\ref{Fig2}.b. b) Impact of tumour heterogeneity on invading structures. (Top) Increase in number of fingers per stronger cell with respect to $\mu_{f}$. (Bottom) Increase in average finger length with respect to $\mu_{f}$. Circled numbers relate to results in Fig.~\ref{Fig2}.a.\label{Fig3}}
\vspace{-10pt}
\end{figure}
Fig.~\ref{Fig2}.a illustrates the typical spatial arrangements between stronger and bulk tumour cells in protrusive fingers, in which stronger cells are commonly seen initiating and leading fingers (see movie 6 in supplementary materials).
To quantify these morphological features, a graph-based method is used to identify the contact network of cells leaving the tumour. 
Fingers are defined as groups connected to the tumour body and more than two cells away from the tumour boundary. 
The tip is defined as the cell of the finger which is topologically the further away from the original tumour boundary. 
The length of the finger is the shortest path length between its tip and the original tumour boundary, measured in number of cells. 
Fig.~\ref{Fig3}.a shows the proportion of fingers led by a stronger cell, pooling all data from experiments with 6, 12 and 24 stronger cells. 
In the regime of collective invasion ($ \mu_{c} \leq \mu_b \leq \mu_{s}$), up to 70\% of the fingers can be led by stronger cells, far more than what would be expected from their density at the tumour boundary (about 25-30\%) in average. 
When $\mu_b > \mu_s$, fingers become a transient feature in the invasion and are equally likely to be led by any cell type. 
Consequently, the proportion of fingers led by a stronger cell decreases.
Similarly, when $\mu_b < \mu_c$, bulk cells cannot migrate on their own and any invading structure forming would be pulled by stronger cells. 
The fact that the proportion of finger led by a stronger cell decreases with $\mu_f$ beyond the optimum is however more surprising (see marks 5 and 6 in Fig.~\ref{Fig3}.a).
Why would an increase in their motility reduce their ability to lead fingers?
To interpret this, we plotted on Fig.~\ref{Fig3}.b the number of fingers per strong cell as a function of $\mu_f$ for $\mu_b=0.150$. 
As $\mu_f$ increases, the proportion of strong cells that generate fingers increases to a plateau value of approximately one in three. 
This plateau is expected since cell orientations are not biased in the model and only a finite fraction of them would migrate away from the tumour from the beginning of the simulation. 
The finger initiation step is therefore not responsible for the invasiveness decay at large $\mu_f$.
However, the finger length has a maximum for an intermediate value of $\mu_f\approx 0.350$, consistent with the optimum observed in Fig.~\ref{Fig2}.b.
Sustaining the growth of a finger is therefore distinct from the ability to initiate it. 
Both properties are however complementary to control tumour invasivity; by multiplying the maps of finger number per stronger cell and relative increase in finger length (\textit{cf.} Fig.4 in supplementary materials), we recapitulate the profile of Fig.~\ref{Fig2}.a.
What makes fingers shorter for large $\mu_f$ is the fact that strong cells tend to escape individually soon after fingers are initiated, leaving behind short fingers that are unable to invade further into the tissue (\textit{cf.} movie 6 in supplementary materials).\

Overall, heterogeneities in the migratory properties of cells combined with simple principles of collective motion appear to be sufficient to reproduce a large number of experimentally observed invasion morphologies.\ 
The model shows that strongly invasive behaviours involve a cooperative process between different sub-populations of tumour cells.\
Pattern of cell types along invasive fingers emerge without any requirement for specific interactions between these cell types.
Therefore, understanding the role of population heterogeneity goes beyond characterising the distribution of their biophysical properties;
in particular, one cannot assess the invasive and metastatic potential of a tumour only based on the proportion and strength of its most invasive cells \cite{Vanharanta2013}.\
However, the local tumour environment and dynamic coupling that exists between cells within the population do control morphologies and malignancy. 
This paper provides an illustration of how diverse the emerging patterns of invasion can be with only three cell populations. 
A general approach for the physics of heterogeneities in collective systems is still lacking, but there is no doubt that it would provide an important contribution to understanding a number of biological processes both in animal development and clinical research.

\begin{acknowledgments}
We would like to thank J-M. Di Meglio, F. Graner, P. Hersen, B. Sorre, R.J. Adams and G. Charras for their valuable comments. A.H. would like to thank the University of Cambridge for the award of an Oliver Gatty Studentship in Biophysics. 
\end{acknowledgments}

\end{document}